\documentclass[10pt]{revtex4}
\usepackage{graphicx,color}
\usepackage{bm}
\newcommand{\beq}{\begin{equation}}
\newcommand{\eeq}{\end{equation}}

\begin{document}


\title{Gene--gene cooperativity in small networks}
\date{\today}
\author{Aleksandra M. Walczak}
\affiliation{Princeton Center for Theoretical Science, Princeton University, Princeton, NJ 08544}
\affiliation{e-mail: awalczak@princeton.edu}
\author{Peter G. Wolynes}
\affiliation{Department of Physics and Department of Chemistry and Biochemistry,
University of California, San Diego, La Jolla, CA 92093}
\affiliation{e-mail: pwolynes@ucsd.edu}




\begin{abstract}
 \begin{center}
\noindent \bf {Abstract}\\
\end{center}
We show how to construct a reduced description of interacting genes in noisy, small regulatory networks using coupled binary ``spin" variables. Treating both the protein number and gene expression state variables stochastically and on equal footing we propose a mapping which connects the molecular level description of networks to the binary representation. We construct a phase diagram indicating when genes can be considered to be independent and when the coupling between them cannot be neglected leading to synchrony or correlations. We find that an appropriately mapped boolean description reproduces the probabilities of gene expression states of the full stochastic system very well and can be transfered to examples of self-regulatory systems with a larger number of gene copies.
\end{abstract} 

\maketitle

\subsection*{Introduction}

Systems with multiple copies of the same type of gene interacting in a single regulatory network are encountered both in synthetic systems in the laboratory \cite{Suel,hasty2}, and in natural organisms, especially when sections of their genomes are duplicated during evolution \cite{Kellis}. The small numbers of the molecules of a given type of protein or nucleic acid taking part in gene regulation are a recognized source of stochasticity in gene expression. In elegant work, deliberately increasing the number of copies of genes expressing proteins has been used to decrease noise \cite{Suel, Suel2,hasty2}. Introducing multiple gene copies should, at first sight,  decrease the fluctuations in the gene expression state. Yet the genes' promoter sites must also compete for binding of the transcription factor molecules in the system. In this way, the limited copies of regulatory proteins may introduce a new source of noise. In this paper, we discuss the steady states of systems of genes with protein mediated interactions and show how the stochastic effects can result in additional cooperativity between genes. 

Gene regulatory systems can be described at varying levels of detail. Genes in large scale networks are usually described as either being expressed \cite{Kauffman}, or not expressed in given experimental conditions. More quantitative understanding of small gene regulatory systems usually requires explicitly keeping track of the number of regulatory proteins, which in turn control the gene expression levels \cite{elowitzleibler, GardnerCollins, hasty1, OzbudakThattai}. Such a detailed molecular description can be formulated in terms of the underlying stochastic processes \cite{sasai, KeplerElston, tenwolde1, tenwolde2}. Buchler, Gerland and Hwa \cite{UliHwa} have argued that transcriptional regulatory systems can be described as Boltzmann machines. Guided by the example of small networks with multiple gene copies, we propose a mapping between a reduced description of a gene network in terms of binary $on-off$ variables and a full stochastic molecular description of regulatory gene-protein systems accounting for protein numbers. This effective two state description links expression states of genes to magnetic networks, which are reminiscent of Boltzmann machines. In this way the present framework connects the description of gene expression states used commonly by cell biologists and the stochastic treatment of the molecular kinetic system.

For illustration we explicitly investigate a set of toy systems, in which we have only one species of protein and multiple gene copies regulated by the given type of transcription factor protein. We consider examples which have known regulatory functions that could be subunits of a larger network - small networks of self-activated and self-repressed genes producing and being regulated by one type of protein. The copies of the gene can be precisely identical copies- that is we model them as having exactly the same chemical parameters; or they can be thought of as mutated versions of the same gene- with different parameters, modeling a system caught in a ``missing-link state" of molecular evolution. We treat both proteomic atmosphere and gene expression states stochastically and use a joint probability distribution, which describes the number of protein molecules in the system (considered to be well-mixed) and the gene expression state to describe each gene [3]. 

\begin{figure}
\includegraphics[scale=0.65]{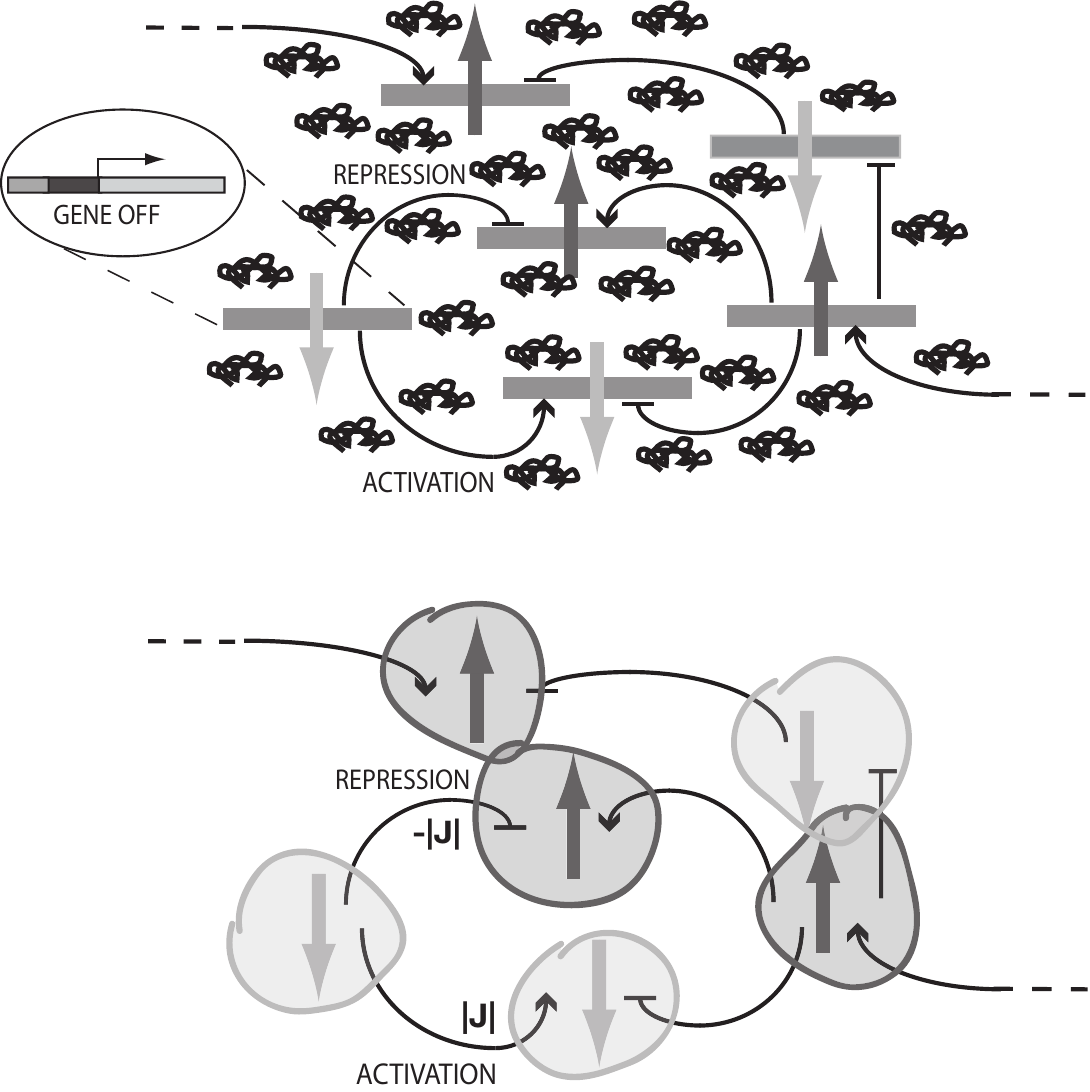}
\caption{A schematic diagram of the mapping of a molecular regulatory gene network onto a two state system. The up and down arrows indicate genes that are respectively found in the $on$ and $off$ states.  The influence of the proteomic field on the gene states is described using local $h$ fields, which describe the genes' tendency to be found in a given expression state, and $J$ couplings, which account for gene--gene interactions. }
\label{Figure0}
\end{figure}

The phenomena described in this paper are examples of collective behaviour. We ask how a switching event in one gene influences the switching behaviour of another gene, to which it is coupled by a mutual protein environment (Figure \ref{Figure0}). For a two body system, the influence of one subsystem on the other may be characterized as a coupling parameter. Strongly interacting elements of a many body systems may lose their individual properties and take on the characteristics of a group acting as a unit in synchrony. For example magnetic spins may couple either directly (``through space"), or indirectly (``through atomic bonds") to form domains of magnetization. It is interesting to ask, whether genes in cells that have a common task, that is produce the same type of protein, actually demonstrate such collective behaviour. We investigate the effects of noise arising from the small numbers of protein molecules and slow gene expression state changes on the strength of cooperativity between genes induced by interaction between the genes and mediating transcription factor molecules.

Using a maximum entropy technique we map the marginal gene expression state probabilities onto those for coupled binary variables in an effective gene field. The sign and magnitude of the effective gene field describes the given gene's tendency to be in a particular expression state. The predicted expression state of the gene can be modified by protein mediated gene-gene interactions which are quantified by the coupling constant. This approach allows us to describe the parameter regimes, in which effective gene expression units can be treated as independent and the parameter regimes, where, in contrast, genes form a strongly coupled unit in the steady state.

We find that for parameter regimes, in which all the genes individually can maintain their own proteomic field and be expressed at an enhanced level, these genes can be treated as independent units. Genes are said to cooperate when at least one of the genes alone could not sustain a proteomic field needed by that gene to produce protein molecules at an enhanced level. In this case, both genes use the mutual reservoir of proteins, and accordingly modify their expression levels. The coupling between the genes results in steady state probability distribution that could not be observed in an effective single gene system. The parameter regime in which genes can be treated as independent is analogous to the case of dressed effective particles in traditional many body condensed matter systems. The properties such as mass of the particle can be modified, but the effective particles can still be treated as independent. In this analogy, the cooperative region is akin to that of strongly correlated electron systems. For potentially bistable systems, we show that the parameter regimes in which the genes cooperate correspond to bimodal probability distributions. In the limit of strongly non-adiabatic binding of transcription function proteins, the gene field clearly depends on the binding and unbinding rates of the gene, whereas the coupling constant is linked to the number of molecules in the proteomic field. 

We test the validity of the transferability of the deduced boolean approximation on a four gene system by comparing the marginal probabilities based on the maximum entropy model to probabilities from simulation studies. We show that even though there are several gene copies, concatenating the two gene interactions accounts for a majority of the correlation in such a system and the resulting description is capable of accurately reproducing the exact marginal probabilities.

\subsection*{The mapping}
\begin{figure}
\includegraphics{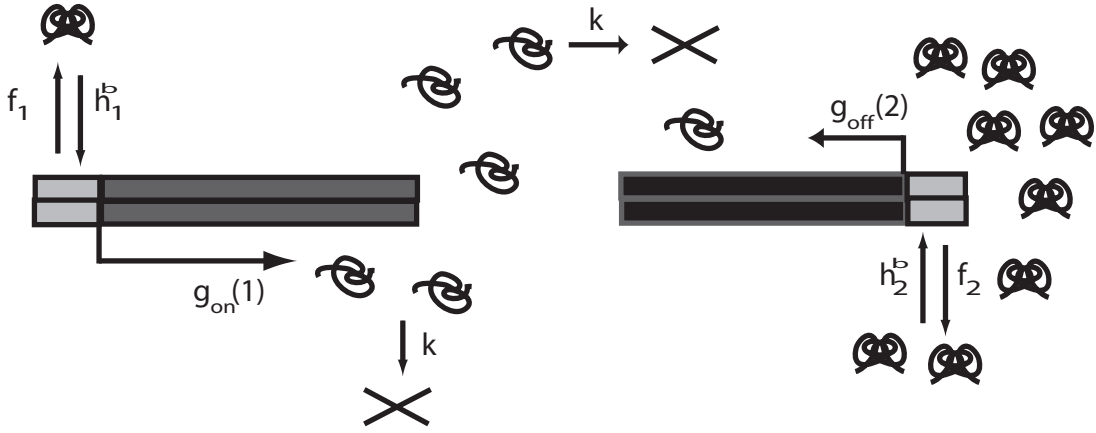}
\caption{A schematic diagram of an interacting gene network with one type of protein and the considered reactions.}
\label{Figure1}
\end{figure}

In order to stochastically describe a set of  $N$ genes which are regulated by binding and unbinding of the same protein, we consider the set of equations for the joint probability distribution $P_{j_1 ...j_N}(n,t)$. This $2^N$ state function describes the state of probability of finding each given gene $i$ is a state $j_i={on, off}$ and the number of regulating protein molecules present in the cell. Since we are only considering one type of protein in these problems, there is a single protein number variable in the master equation. The state of the gene switches from one for which there is only basal production (the gene is $off$) to another state of activated production (the gene is $on$) by binding/unbinding of activator/repressor proteins. To be explicit, for example a two gene system is described by a four state probability vector: $[P_{on,on}(n),P_{off,on}(n),P_{on,off}(n),P_{off,off}(n)] $. The number of protein molecules in each state can change due to production of proteins by a given gene $i$ in a given regulatory state with a constant rate $g_{i, on}$, $g_{i, off}$, and degradation, with a rate proportional to the concentration $k n$.  Proteins bind as dimers to the operator sites on genes with a bimolecular rate coefficient $h^b_i$ and unbind with a fixed rate $f_i$. The processes considered in the problem are depicted in Figure \ref{Figure1}. The evolution of the whole system is described by a $2^N$ dimensional matrix coupled master equation. The gene expression states are coupled by binding and unbinding of proteins. We present detailed forms of the master equation for two activated genes in Appendix A. We solve the full stochastic steady state master equation both by exact numerical methods and by simulating with a random time Monte Carlo (Gillespie) algorithm \cite{gillespie}.

In parallel with the full description, we can introduce a reduced description of the gene expression states specifying only probabilities of the genes being $on$, regardless of the number of proteins in the system $C_{j_1j_2 ...j_N}=\sum_{n} P_{j_1j_2 ...j_N} (n)$. The relation is easily generalized to a system with $K$ types of proteins to $C_{j_1j_2 ...j_N}=\sum_{n_1}..\sum_{n_K} P_{j_1j_2 ...j_N} (n_1,.. n_K)$. Mapping the state of each gene onto a discrete spin-like variable $s_i={0,1}$, these reduced probabilities can be rewritten in terms of an Ising model with site specific fields $h_i$ and coupling constants $J_{ij}$, as depicted in Figure \ref{Figure0}:
\beq
C_{s_1s_2 ...s_N}=\frac{1}{Z} \exp[(\sum_{i} h_i s_i+\sum_{i<j} J_{ij} s_i s_j) ]
\eeq
where $Z=\sum_{\{s_i\}} \exp[(\sum_{i} h_i s_i+\sum_{i<j} J_{ij} s_i s_j)$. Although we focus here on a reduced description using marginal probabilities of the gene states and protein molecule number independent parameters ($h_i$ and $J_{ij}$), in general the same relations hold for $n$ dependent fields $h_i(n)$ and coupling constants $J_{ij}(n)$. The marginal probabilities $C_{ij}$ are replaced by the full protein dependent probabilites $P_{ij} (n)$ in the expressions for the parameters, and the fields and couplings must be evaluated for each protein number state separately. 

We have chosen to use the nonsymmetric from of the Ising model Hamiltonian having spins $s_i=\{0,1\}$, as opposed to the more usual choice in solid state physics that employs a symmetric model with $s_i=\{-1, +1\}$. This choice is motivated by the lack of symmetry in the interactions between the $on$ and $off$ states. The genes communicate with each other by protein mediated interactions, which increase with the protein concentration. When both genes are in the $off$ state, the protein concentration is low, so it is not natural to regard the genes as ``interacting" to stabilize the $off, off$ state, as the symmetric Ising would imply. In an intuitive sense, only the $on, on$ state can be stabilized by the protein field (or destabilized in the case of a repressor), hence it requires an interaction term. When one gene is $on$ and the other $off$ the protein field in which both genes function is not strong enough to stabilize the $on, on$ state, again the genes effectively do not interact. We can see that the biochemical details of the interplay of the binding processes make it more convenient to use the nonsymmetric formulation of the Ising model. For concreteness we summarize the mapping onto a symmetric model along with the main differences in the interpretation of the fields and interactions in the symmetric and nonsymmetric representations in Appendix B.

For a two gene system, there is a direct mapping between the marginal probabilities of the genes and the Ising model parameters. Explicitly,
\begin{eqnarray}
h_1=\log \frac{C_{on, off}}{C_{off, off}}\nonumber \\
h_2=\log \frac{C_{off, on}}{C_{off, off}}
\label{hitwoU1}
\end{eqnarray}
\beq
J_{12}=\log \frac{C_{on, on} C_{off, off}}{C_{off, on}C_{on, off}}
\label{hitwoU2}
\eeq
The external field $h_i$ describes each of the gene's tendency to be in the $on$ state (or $off$ state for negative values), by comparing its probablity to be $on$ when all the other genes are $off$ to the case when all genes are $off$. Positive values of $h_i$ indicate the gene is more likely to be in the activated state. The coupling constant expresses quantitatively the cooperativity of two genes in the system, i.e. whether they will be simultaneously found to be active or not. Large absolute values of the coupling constant indicate a high probability for two genes to be simultaneously $on$ compared to the genes being $on$ independently in an uncorrelated fashion. Positive couplings reinforce the proteomic field. Negative couplings inhibit the effect of the proteomic field. For cooperative genes the actual states of the genes in a given region of parameter space that are observed depend on the coupling constants.

A key quantity compares the rate of fluctuations in the gene's occupancy state to the rate at which the protein number changes by synthesis or degradation. We call this ratio the adiabaticity parameter, which compares the characteristic timescales for the two processes. To estimate the timescale associated with the change of the gene expression state we compare the binding time to the protein lifetime $\kappa_i=\frac{h^b_i g^2_{i on}}{k^3}$. Small adiabaticity parameters describe gene occupancy states which change on long timescales compared to timescales on which protein numbers change. Large values of the adiabaticity parameter indicate that the gene occupancy equilibrates before the protein numbers reach a steady state. This is the limit usually considered in the so-called thermodynamic framework, such as the one originally proposed by Shea and Ackers \cite{SheaAckers} and utilized by Buchler et al \cite{UliHwa} in their Boltzmann machine analogy.

We expect the external field $h_i$, better termed the ``internal field" as it describes the gene's tendency to be in the $on$ state, to be related to the equilibrium binding constant of the gene site, and to the protein concentration. Despite a general complicated nonlinear expression for the $h_i$ field of each gene, the leading order term is dominated by $h_i=\log((\frac{2<n>}{n_i^+})^2)=\log(\frac{h^b_i <n>^2}{f_i})$. In the adiabatic limit the mean protein number $n$ can be determined from the deterministic equations of motion, $n=\sum_i \frac{g_{i1}h^b_i n^2+g_{i0}f_i}{h^b_i +f_i}$ for a set of activated genes.  In the nonadiabatic limit, the mean number of proteins is determined by the sum of the production rates of the genes in the state with the largest marginal probability. We will discuss the two limits in greater detail throughout the paper. 
%

\begin{figure}
\begin{center}
\includegraphics{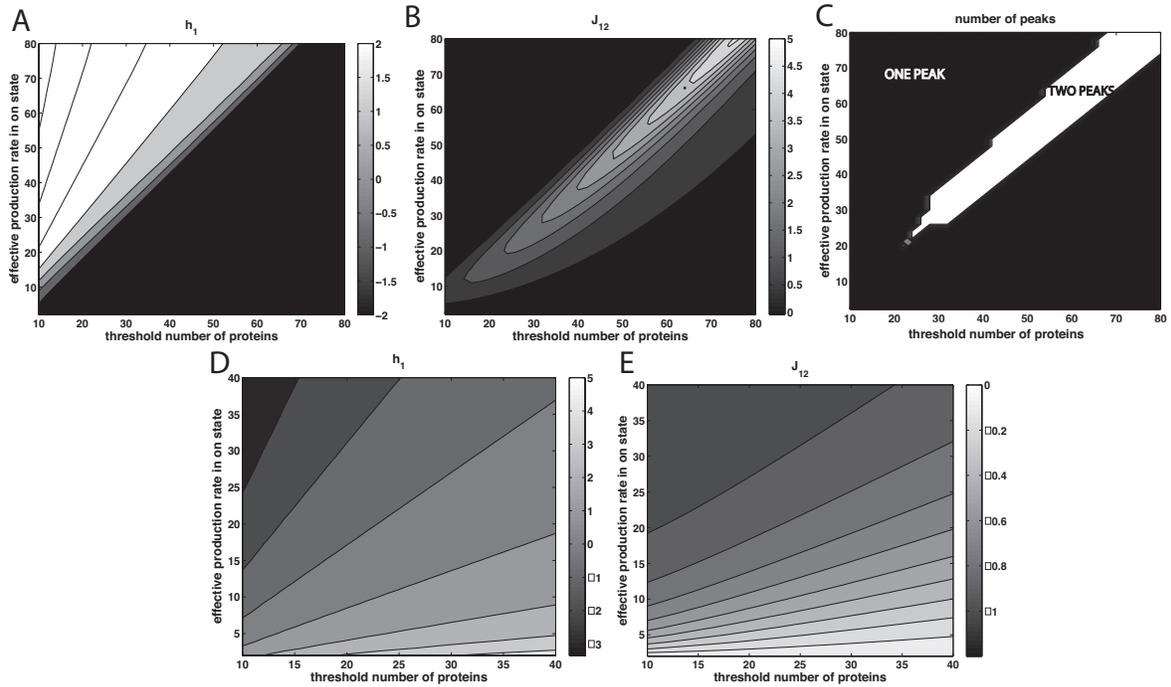}
\caption{The gene $h$ fields (\textbf{A}), couplings $J_{12}$ (\textbf{B}) and number of peaks (\textbf{C}) in the distribution for a system with two symmetric adiabatic activated genes as a function of the effective production rate in the $on$ state $\frac{g_{on}(1)+g_{on}(2)}{k}$ and the threshold number of protein molecules $n^{\dagger}$. The parameter of both genes are varied. The fixed parameters are: $k=2$, $g_{off}(2)=2$,$\kappa=720$. The gene $h$ fields (\textbf{D}) and couplings $J_{12}$ (\textbf{E}) for a system with two weakly adiabatic repressed genes as a function of the effective production rate in the $on$ state $\frac{g_{on}(1)+g_{on}(2)}{k}$ and the threshold number of protein molecules $n^{\dagger}$. The parameter of both genes are varied. The fixed parameters are: $k=2$, $g_{off}(2)=2$, $\kappa=1$.}
\label{Fig3}
\end{center}
\end{figure}

\subsection*{Two copies of the same gene}
\textbf{The adiabatic regime}\\
\textbf{Two activated genes}\\
We first consider the case where the gene expression state can safely be taken to be in equilibrium with the proteomic atmosphere. Using deterministic rate equations, a system of two copies of an identical gene (with identical equilibrium constants) can be described as a composite one gene system having $<n>=\frac{g_{on}^{eff} h^b n^2+ g_{off}^{eff} f}{ h^b n^2+f}$, with a doubled production rate $g_{on}^{eff}=2 g_{on}$, $g_{off}^{eff}=2 g_{off}$. A more detailed full stochastic analysis reveals differences in the probability distributions of the composite single effective gene and the actual two gene system in certain areas of parameter space. Through their correlations two identical copies of the genes interact and their cooperativity modifies the mean number of protein molecules actually found in the system.

In order to understand the interplay between the two genes, it helps to consider a phase diagram for the parameters of the reduced description in terms of $on$/$off$ gene expression state variables. Figures \ref{Fig3} $A, B$ shows a phase diagram for the gene $h$ field and coupling constant $J_{12}$ of the two genes, with different equilibrium parameters as a function of the effective production rate of the whole system ($n^{eff}=(g_{on}(1)+g_{on}(2))/k$) and the threshold number of proteins at which each gene alone would be equally likely to be $on$ and $off$. The positiveness of the $h$ field describes each gene's specific tendency to be found  in the $on$ state, in a certain parameter regime. As $n^+$ increases, more proteins are needed to turn $on$ each of the genes. For larger $n^+$ the genes are more likely to be $off$ for larger production rates, which is quantified by an increasing negative $h$ field. The $h$ fields of the genes describe effective independent genes dressed by the proteomic field. Whether the genes actually will be on, depends on the coupling quantifying the correlations between the two genes. If the $h$ field is large and positive, each of the genes produces a proteomic field that sustains its own $on$ state. For smaller values of the $h$ field the tendencies of the gene to be $on$ and $off$ are similar and the observed gene states depend on the genes' cooperativity -  described by the coupling constant. For a system having only activated genes leading to interactions the $J_{ij}$ coupling is always positive. It is favourable for both gene copies to be in the $on$ state, since they change their state by responding to a mutually available proteomic field. The positive $J_{ij}$ coupling, stabilizes the $on, on$ state and increases the probability of a given gene to be $on$. Therefore genes cooperate when they have a slightly higher probability to be $on$, rather than $off$.

The cooperation results in a sharper transition as a function of average protein molecule numbers $n$ from the $on$ to $off$ state, than if the genes could be assumed to be strictly independent in their binding events. Such a cooperativity  between genes is a result of coupled genes and is a second order effect compared to the direct response to the available proteomic field. The genes cooperate to sustain a collective proteomic field. From the study of a single activated gene, we know that self-activated genes become bistable when the probability to be $on$ is similar to that for the gene to be $off$.  We can see that for systems with two identical activated genes the region of interacting genes ($J>0$ in Figure \ref{Fig3} $B$) correspond to regions of bistability (\ref{Fig3} $C$) of the solutions of the two gene system. Therefore when studying the attractor structure of large networks we are intrinsically focussed on the regions of phase space when genes are strongly interacting.

\textbf{Two repressed genes}

The difference between the behaviour of systems made up purely of gene activators and those having purely repressors can already be seen within the deterministic equations, where the repressed gene system always exhibits only one stable steady state solution. This solution corresponds to a monostable probability distribution in the adiabatic regime. In the case of an activated gene, if the effective production rate ($n^{eff}$) is larger than the threshold number of proteins the gene will be $on$ and the number of proteins is governed by the balance of the production and degradation rates. In the case of a system with a repressed gene, however, very high production rates in the on state will result in repression of the gene. 

A system of two symmetric genes with repressors in the adiabatic regime can deterministically always be well described by one effective repressed gene. As in the case of two activated genes, the proteins produced by both genes are shared by both genes as repressive transcription factors. At the same time, unlike the case having symmetric activated genes, the repressed genes are more strongly coupled throughout all of the parameter range (\ref{Fig3} $E$). Much like what is seen for the system with activators however, the couplings play a role when the $h$ gene field changes sign. In that region of parameter space the genes have a slightly larger probability to be $on$ than $off$. Figure \ref{Fig3} $D, E$ show an example of a symmetric system of two repressed genes in the weakly adiabatic regime. Here the region corresponding to significant correlation between the genes is larger than that found in the strictly adiabatic regime. Increasing the production rates in the $on$ state, destabilizes the $on$ state (decreases the gene field $h$), by increasing the binding rates of repressors. This in turn results in small numbers of protein molecules in the system. The decrease in the gene field $h$ with the increase of the effective production rate is a result of negative feedback in the repressed gene system, as opposed to the positive feedback in the case of the activated gene, which results in an increase of $h$. Since the genes are coupled by proteins, and increasing protein numbers destabilizes the $on$ states, a gene found in the $on$ state induces the other gene to be in the $off$ state. Therefore genes are hence less likely to be found in the $on, on$ state compared to the $on,off$ or $off,on$ states and the effective binary couplings for repressed genes have negative signs. The probability of a gene being in the $on$ state is reduced compared to a noninteracting system. Negative couplings reduce the number of protein molecules, and effectively stabilize the $on$ state. If the intrinsic preference of the genes to be $on$ is small (positive $h$ gene fields close to zero), the negative coupling destabilizes the $on$ states, and the genes are more probable to be found in the $off$ state. When the thresholds are small compared to protein molecule numbers produced in the $on$ state, the gene is repressed regardless of the production rate. In this region of parameter space the number of protein molecules is dominated by the effective production rate in the $off$ state. Since for low threshold numbers and high effective production rates switching events are likely, the couplings are large and depend mainly on the threshold number of protein molecules, as those regulate the number of protein molecules. Only for small effective production rates and large threshold numbers is the average number of protein molecules small without additional feedback from a repressed gene and the coupling constant tends to zero.

Since the steady state probability distribution for the adiabatic repressed gene system has one peak, we can further explore the analogy between the Poisson description of the problem and the couplings of the pseudo-spin system. We expect the probability of the number of protein molecules for a system with a single adiabatic gene to be described by a Poisson distribution. We can link negative $J_{12}$ couplings with a decrease in the width of the probability distribution below that expected from the Poissonian case $<n^2>-<n>^2<<n>$. The decrease in the spread of the distribution compared to the Poisson case, also holds true when the system is treated as an the effective one gene system, where the mean number of proteins is given by $<n>=\frac{g_{off}(eff) (\frac{<n^2>}{n^+})^2+g_{on}(eff)}{1+(\frac{<n^2>}{n^+})^2}$. Therefore these sub-poissonian statistics are not associated in a simple way with increasing the number of genes. The observed statistics are a result of the constraining interactions for self-repressing systems, which decrease the variance of the protein molecules. However in parameter regions where couplings cannot be neglected, the variance is slightly smaller for the two gene case compared to that for the composite one gene description. Adding more gene copies, enforces the constraints on the spread of the number of protein molecules, by introducing additional repression from the other genes. The decrease in noise in a many repressed gene system is therefore not a result of averaging the gene occupancy states. In the adiabatic limit, the gene occupancy state is equilibrated and adding more genes should not modify the gene occupancy statistics. The decrease in the spread of the distribution is a result of constraining protein molecule number fluctuations by introducing additional control mechanisms in the form of other repressed genes. This control is implemented as the feedback of collective proteomic field. For the activated system, the positive couplings can be related to an increase of the variance compared to the mean for protein molecule numbers $<n^2>-<n>^2><n>$. In this case however, the increase of the spread of the number of protein molecules is not related to broadening of the peak of the distribution, but to the emergence of two peaks and a bistable distribution. The value of the coupling parameter, both for the repressed and the activated systems, is a measure of departure from Poisson statistics. 

\begin{figure}
\begin{center}
\includegraphics{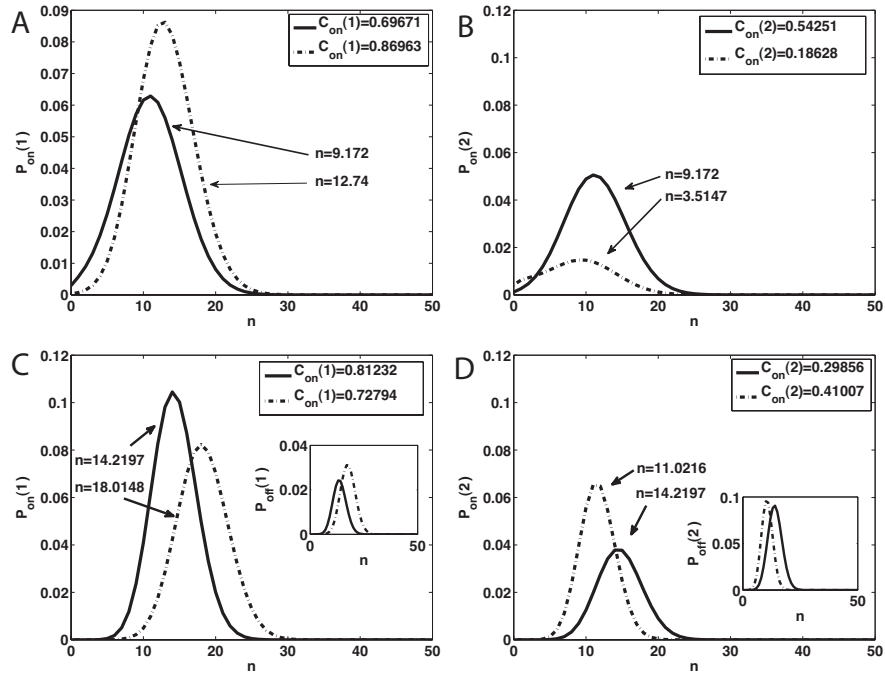}
\caption{The probability distributions of the number of protein molecules and the gene to be found in the $on$ state for gene $1$ (\textbf{A}) and gene $2$ (\textbf{B}) in a system with two asymmetric adiabatic activated genes (solid lines) compared to the probability distributions of an effective single gene with the same effective production rates as the two gene system. $C_{on}(i)=\sum_n P_{on}(i)$ is the probability of gene $i$ to be in the $on$ state. $g_{on}(1)=40, g_{off}(1)=1,g_{on}(2)=17, g_{off}(2)=1, n^+_1=3, n^+_2=8$ and $k=720$. The probability distributions of the number of protein molecules and the gene to be found in the  $on$ state for gene $1$ (\textbf{C}) and gene $2$ (\textbf{D}) in a system with two asymmetric adiabatic repressed genes compared to the probability distributions of an effective single gene with the same effective production rates as the two gene system. The inset shows the probability distributions of the number of protein molecules for the gene to be found in the $off$ state.$g_{on}(1)=12, g_{off}(1)=1,g_{on}(2)=12, g_{off}(2)=1, n^+_1=30, n^+_2=10$ and $k=720$.}
\label{Fig4}
\end{center}
\end{figure}

\begin{figure}
\begin{center}
\includegraphics{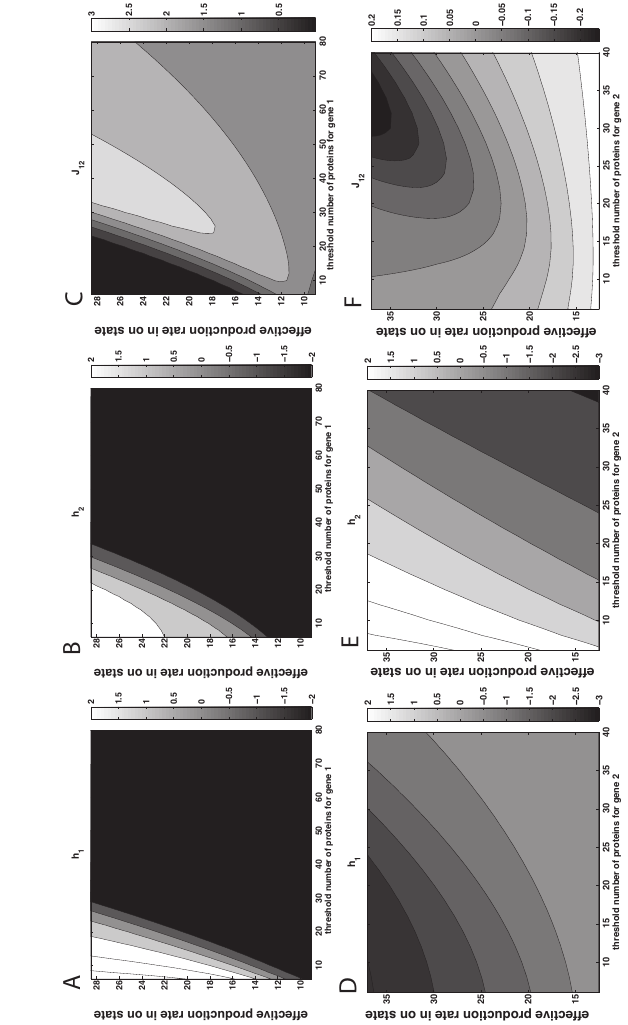}
{\linespread{1.0} 
\caption{ The gene $h$ fields for the activated gene $1$ ({\textbf{A}}) and the activated gene $2$  ({\textbf{B}}) and couplings ({\textbf{C}}) as a function of the effective production rate of the system in the $on$ state $\frac{g_{on}(1)+g_{on}(2)}{k}$ and the threshold number of protein molecules $n^{\dagger}_1$ of gene $1$ for a system with two asymmetric adiabatic activated genes. The parameters for gene $1$ are varied, while those for gene $2$ are fixed. The parameters: $g_{on}(2)=17, g_{off}(2)=1, n^+_2=8$ and $k=1$, $\kappa=720$. The gene $h$ fields for the repressed gene $1$ ({\textbf{D}}) and the activated gene $2$ ({\textbf{E}}) and couplings {\textbf{F}}) as a function of the effective production rate in the $on$ state of the system $\frac{g_{on}(1)+g_{on}(2)}{k}$ and the threshold number of protein molecules $n^{\dagger}_2$ of gene $2$ for a system with one repressed adiabatic and one activated adiabatic gene. The parameter of the activated gene (gene $2$) are varied. The fixed parameters are: $g_{on}(1)=24, g_{off}(1)=2, n^+_1=8$ and $k=2$, $g_{off}(2)=2$. }
\label{Fig16}}
\end{center}
\end{figure}


\textbf{Two different activated genes}

We now consider two similar genes that while having different binding rates bind and produce the same protein. In this case there is no ``composite" single gene that will reproduce the steady state distribution of protein molecules.  In Figures \ref{Fig4} $A, B$ we compare the two gene system to two separate one gene systems where each of the genes has the same effective production rate as the two gene system. The binding and unbinding parameters in the one gene systems are the same as those of the individual genes in the two gene systems. We see that in the two gene system the steady state probability distributions reflect the cooperation between the two genes. The gene with the weaker promoter, alone does not have the ability to stabilize a proteomic cloud that would keep the gene in the $on$ state (activator bound) (dashed line in Figure \ref{Fig4} $B$). When the genes cooperate, the gene having the weaker promoter, makes use of the reservoir of proteins sustained by the gene having the stronger promoter and therefore the weakly promoted gene is now more likely to be found in the $on$ state (solid line in Figure \ref{Fig4} $B$). Such a steady state could not have been sustained by the weak promoter gene alone without the external help of the more strongly promoted gene. The cooperation between genes entails the genes sharing their mutual proteomic reservoir. As a result of sharing the available protein molecules, the gene with the stronger promoter has access to a smaller number of activators and its probability to be on decreases (solid line in Figure \ref{Fig4} $A$) compared to what would be found for the effective one gene system (dashed line in Figure \ref{Fig4} $A$). As can be seen from the example in Figures \ref{Fig4} $A, B$ the overall mean number of proteins slightly decreases in this cooperative system, when available resources are distributed between the two genes. The example presented in Figures \ref{Fig4} $A, B$  suggests, that in a system with initially two identical copies of a gene with strong promoters, under the wing of one gene which retains a strong promoter and sustains the proteomic field, the other copy could evolve to have completely different binding characteristics than the two genes had originally. 

In the example just described, the occupancy states of the individual genes are regulated by the strengths of their promoters, and are therefore different. The phase diagrams presented in Figure \ref{Fig16} $A, B, C$ shows the gene $h$ fields and coupling constant $J_{12}$ of the two genes with different equilibrium parameters, as a function of the effective production rate of the whole system and the threshold number of proteins at which gene $1$ alone would be equally likely to be $on$ and $off$.  In these phase diagrams we vary the parameters of the first gene -- the one that was more probable to be found in the on state. As $n_1^+$ increases, more proteins are needed to turn gene $1$ $on$, and the gene has a tendency to be $off$ for larger production rates, which is reflected in an increased negative $h_1$ field. The other gene, which alone would be $off$, reacts to any extra proteins coming from the first gene, by increasing its tendency to be $on$, quantified by the positive $h_2$ field. However, as the threshold $n^+_1$ of  gene $1$ increases, gene $1$ becomes less probable to be $on$, and hence less able to sustain a protein reservoir. As the number of proteins in the proteomic reservoir decreases, the second gene decreases its tendency to be on. Whether the genes actually will be on depends on the tendencies of both genes. If both of them have positive $h$ fields, they can easily produce a proteomic field that sustains the two $on$ states. If one of the genes has a negative $h$ field, the observed gene states depend on the gene cooperativity - the coupling constant. The positive $J_{ij}$ coupling increases the probability of the weak promoter gene to be $on$ and reinforces the $on-on$ state, as we saw in the example in Figures \ref{Fig4} $A, B$. From the phase diagram,we see the genes interact when one gene has a positive $h$ gene field and the other a negative $h$ gene field.

Compared to two copies of identical genes we can quantitatively see that different genes cooperate with each other throughout a much greater region of parameter space. In the cooperative region the protein field produced by one gene is not enough to sustain the enhanced production by that gene. However, the cooperation of the identical genes is stronger (larger values of $J_{12}$). A system with two different genes having only activators can be bistable, in a region of parameter space where the individual genes with production rates adjusted to mimic those of the two gene system would be monostable. Hence gene-gene interactions, encoded by the coupling constant, not only have quantitative effects, but can modify the attractor structure of a gene regulatory system.

\textbf{Two different repressed genes}

We can also study a system of two genes that  produce proteins that act on both genes as repressive transcription factors. In this case, the gene that has a higher threshold number of proteins neeeded to change its occupancy (a gene which will get repressed at larger concentrations -- gene $1$), acts as a buffer (Figures \ref{Fig4} $C, D$). The gene, that provides the buffer (gene $1$), remains in the $on$ state throughout those regions of parameter space where the gene with a lower threshold number of proteins (gene $2$) would have been more likely to be found in the $off$ state. The gene with a lower threshold number of proteins (gene $2$) alone could reach a steady state with an intermediate value of the probability to be $on$. However the buffer gene (gene $1$) provides a reservoir of repressors, which stably represses the buffered gene (gene $2$). The insets explicitly show the probability distributions as a number of protein molecules of the two genes to be found in the $off$ states, since the buffered gene (gene $2$) has a higher probability to be $off$ than to be $on$. The couplings $J$ for the genes presented in this example are negative, indicating that the variance of the two gene distribution is smaller than that of a Poisson distribution.

\textbf{One repressed and one activated two gene system}

A system which consists of a single repressed gene and a single activated gene will jointly show the characteristic behaviour of either an effective single weakly activated gene or an effective weakly repressed gene, depending on the values of chemical parameters. For large numbers of protein molecules in the system the activated gene will be $on$ and the repressed gene will be $off$. Conversely, for small numbers of protein molecules, the repressed gene will be $on$ and the activated gene $off$. For example, when the gene to be activated has a higher threshold, the activated gene is $off$ and the repressed gene $on$. In most of the parameter range, a state with one gene in the $on$ state and the other in the $off$ state is the most probable. In the adiabatic regime, this system will always be monostable, as the activated gene requires strong differences between the $on$ and $off$ production rates for bistability. If we consider a tug of war symmetric repressed-activated two gene system (results not shown), we find it practically does not cooperate (close to zero negative coupling constant), as both $h$ fields completely cancel each other in the $on, on$ state, giving a natural advantage to the $on, off$ states. In such case, the close to zero couplings in an adiabatic system describe effective independent genes and hence result in close to Poissonian probability distributions.

Figures \ref{Fig16} $D, E, F$ show a phase diagram for a repressed-activated nonsymmetric two gene system. We modify the parameters of the activated gene (gene $2$), and keep the repressed gene (gene $1$) in the $off$ state. When the $h$ field of the activated gene balances the $h$ field of the repressed gene, the couplings can be neglected. When the $h$ fields of the two genes are both negative and the field of the repressed gene is small (bottom half of Figures \ref{Fig16} $D, E, F$), the couplings are positive. In the repressed-activated two gene system, if the number of protein molecules is small, the repressed gene becomes unrepressed and the number of proteins increases. An increase in the number of proteins increases the positive coupling which stabilizes the $on,on$ state. Yet despite an increase in the probability for both genes to be found in the $on$ states, compared to independent genes in a similar proteomic field, in the bottom half of Figures \ref{Fig16} $D, E, F$ the threshold values are such that the repressed gene is slightly more likely to be found in the $off$ state, as is the activated gene. If the genes are not symmetric, a state with the repressed and activated gene having the same probability to be $on$ is not a steady state solution. The states with both the repressed and activated gene in the same occupancy state ($on,on$ and $off,off$) are not likely. In the case of two nonsymmetric activated and repressed genes the couplings between genes act to counterbalance the $h$ fields of the two genes. An example of such a case is shown in the top right of Figures \ref{Fig16} $D, E, F$ where the fields are of opposite signs, the activated gene has a small field and the couplings are negative. High threshold values for the activated gene result in the gene being found in the $off$ state. The repressed gene is more likely be found in the $on$ state. The negative couplings show that the $on, on$ state is not a likely not a steady state solution. The $on,on$ state would result in enhanced production by the activated gene, which would lead to an increase of proteins and the repressed gene would be turned $off$.  The strength of the couplings is limited by the number of protein molecules in the system. When the $h$ fields of the two genes have large and opposite values, the couplings become small (top left hand corner of Figures \ref{Fig16} $D, E, F$). In this case, for large production rates and small threshold parameters the system acts as an effective activated gene system. For large threshold parameters and small production rates the two gene system (bottom right corner of Figures \ref{Fig16} $D, E, F$) acts as an effective repressed gene system and for large production rates. In the regime where the effective activated gene dominates, the activated gene produces enough proteins to stably repress the repressed gene and activate itself. Similarly to a pure two activated gene system, the couplings can be neglected.

We can again link the sign of the couplings with the relation of the variance to the mean for the number of protein molecules. The Poisson distribution, which is expected for a noninteracting adiabatic system, has a variance of the number of protein molecules equal to the mean number. In the example in Figure \ref{Fig16} $D, E, F$ the distribution changes from having a variance less than the mean for negative couplings in the activated gene dominant regime to having a variance larger than the mean in the repressed gene dominant regime. Therefore the sign and value of the couplings, which are a result of the interplay between activated and repressed genes, are related to the noise characteristics, specifically the variance of the probability distribution of the number of protein molecules, of a gene functioning in a many gene network. In the self-activating system, the numerical results indicate the couplings have a logarithmic dependence on the variance $J_{12} \sim \log(\alpha(<n^2>-<n>^2)/<n>)$ where $\alpha$ is an unknown constant. We plan to exploring this relation and its consequences.

The repressed-activated two gene system is analogous to an antiferromagnet, since the most favourable configuration for the pseudo-spins representing the genes is to be antialligned. In this mapping of the molecular description onto an analogous Ising model, antiferromagnetism arrises primarily because the gene $h$ fields of the repressed and activated gene are of opposite signs in most of the parameter range. When the genes have $h$ fields of the same sign, the intrinsic frustration is reintroduced by means of the $J_{ij}$ couplings. Only when the protein field is very small, or very large, one gene represses the other and the system loses its balance. Therefore coupling many repressed-activated two gene systems with similar probabilities to be $on$ will lead to frustration.

\textbf{Introducing genes into an existing system}

Consider two alternative ways for introducing an additional copy of a given gene into an existing system having only a single copy. In one manner of doing this we may assume that all other elements of the system remain fixed. One does not control for increasing the steady state number of protein molecules by modifying degradation mechanisms or limiting the number of resources in the cell. Alternatively we can consider a more mathematically convenient limiting case that corresponds to a model which limits the production rate of proteins even when the additional gene is present. The simplest example of limited resources is a limited number of RNA polymerases in the cell, which would limit transcription, despite a potentially large potential rate of mRNA production. Another limiting step could be the lack of aminoacid building blocks, which would in turn block translation. The scenario of an unrestrained increase in mean protein numbers is of limited relevance in the many gene limit (and hence the cell), as lack of protein building blocks, cell machinery and crowding quickly become issues. Especially the fact that there is a very limited number of complicated complexes, such as RNAP, justifies the detailed study of the second scenario, which we have thus far discussed, in which the protein production rate per gene is modified when introducing a new gene. However if the protein production resources are abundant introducing small numbers of new copies of given type of gene into the system can increase the overall protein production.

Let us consider the steady states of a two gene system in which the two genes are coupled by binding and unbinding of the same type of activator protein. Not suprisingly, adding a new activated gene, simply increases the number of protein molecules in the activated-activated gene system compared to the case having a single gene producing that kind of protein. In the limit of fast binding and unbinding of transcription factors, the two gene system, with similar equilibrium constants, may described well by an effective single gene with parameters equal to the sum of the production rates of the two genes in the particular states ($g^{eff}_{on}=g_{1on}+g_{2on}$, $g^{eff}_{off}=g_{1off}+g_{2off}$). Although this relation to a single gene system is not surprising, a qualitatively new behaviour appears: since the number of protein molecules in the system changes, a gene that was previously bistable can now become monostable and vice versa \cite{hat}. Furthermore, since the resulting system characteristics are driven by the mean number of protein molecules, which has increased even a ``composite gene" made up out of genes which would be more probably in the $off$ state can now be completely activated, due to the shift of $K^{eq}=h_i <n^2>/f_i$. 

\subsection*{The nonadiabatic regime}
We now consider systems for which the assumption of very fast binding and unbinding of the transcription factor from the gene expression binding site is not valid. We must then explicitly consider the effects of the change in gene expression state on the time averaged properties of the system \cite{idopaulssonbact, Idorev, WangIdoAustin, Walczak, hornos}.  In this regime the protein number now has time to come to a steady state before the gene occupancy state changes and there are two separate peaks for the on and off state in the probability distribution, as opposed to a single equilibrated peak characteristic of the adiabatic distribution. For this case even systems having two copies of the same genes show strong cooperativity and the genes can neither be described by a single effective gene model nor can they be treated separately as independently fluctuating units. A strong signature of nonadiabaticity is visible in the phase diagrams: the couplings are mainly a function of the synthesis rate, whereas the $h$ gene field now depends mainly on the threshold number of protein molecules.

\begin{figure}
\includegraphics{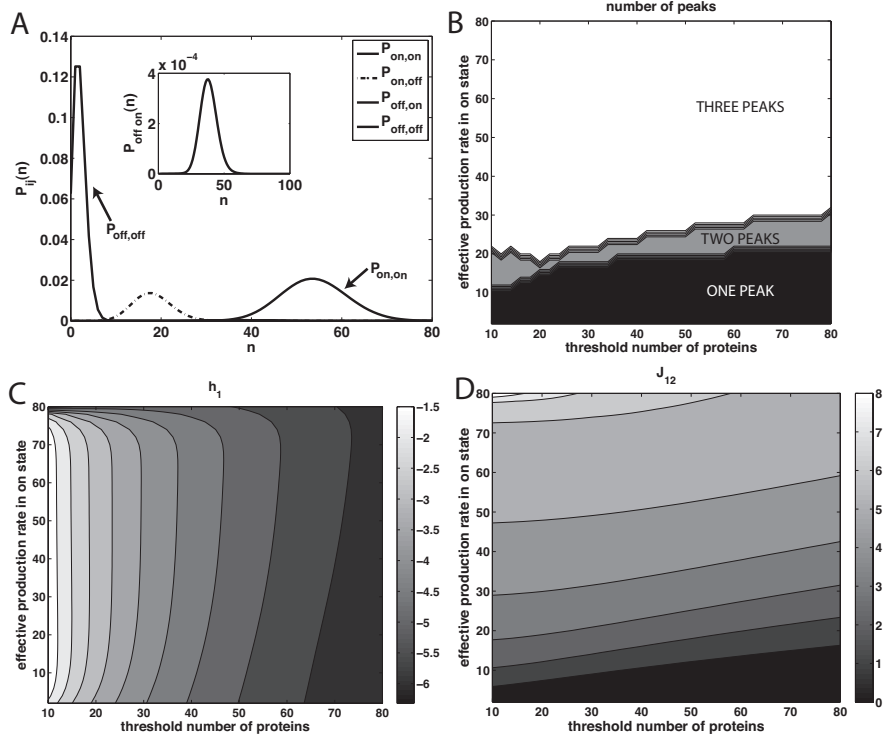}
\caption{\textbf{A}: The four components of the probability distributions of the number of protein molecules for the different gene occupancy states: $\{on,on\}$, $\{on,off\}$, $\{off,on\}$, $\{off,off\}$ of a system with two asymmetric nonadiabatic activated genes. The inset shows the probability distributions of the $off, on$ state separately. $g_{on}(1)=40, g_{off}(1)=1,g_{on}(2)=17, g_{off}(2)=1, n^+_1=3, n^+_2=8$ and $k=720$. The number of peaks in the distribution (\textbf{B}) and the gene $h$ fields (\textbf{C}), couplings (\textbf{D}) as a function of the effective production rate in the $on$ state $\frac{g_{on}(1)+g_{on}(2)}{k}$ and the threshold number of protein molecules $n^{\dagger}$ for a system with two symmetric nonadiabatic activated genes. The parameter of both genes are varied. The fixed parameters are: $k=1$, $g_{off}(2)=1$,$\kappa=0.5$.}
\label{Figure8}
\end{figure}

\textbf{Two activated genes} 
 
 For a system of two activated genes the coupling strengths generally dominate over the $h$ fields (Figure \ref{Figure8}) $C, D$. The coupling constant $J_{12}$ increases with the number of protein molecules available in the proteomic reservoir, because the cooperativity is mediated by the proteins. The cooperativity between the genes increases with the protein concentration. The gene $h$ field depends mainly on the equilibrium constant (related to the threshold number of protein molecules). The $h$ gene field describes each genes own tendency to react to the proteomic field. The resulting gene occupancy state is once again a result of both the gene fields and the coupling between the genes. The probability to be in a occupancy given state depends on both the production rate and the threshold $n^+$. The $h$ gene fields depend mainly on the threshold value while the couplings mainly depend on the production rate. This separation of dependencies is a signature of nonadiabaticity. In the nonadiabatic limit proteins equilibrate in a single gene occupancy state. In the adiabatic limit, proteins equilibrate in an effective gene state, which is reflected by a dependence of the $h$ and $J$ fields on both order parameters.
 
In the nonadiabatic limit the occupancy state of each genes changes primarily according to the binding and unbinding rates of the particular gene. The protein molecule numbers equilibrate in each of the occupancy states. As a result of this slow equilibration there are four possible gene expression states for the composite two gene system. However, since basal production rates are small and the considered enhanced production rates are of the same order of magnitude, effectively the two $on, off$ states are indistinguishable from each other in terms of protein numbers, even if the basal production rates differ. As a result, we find three distinct peaks in the probability distribution for protein numbers (\ref{Figure8} $A$): one which corresponds to both genes producing proteins at the enhanced level, one when both genes produce proteins at the basal level, and the last state, when one gene is $on$ and the other $off$. If we considered a wider range of production rates in the case of a system with two asymmetric genes we could observe four distinct peaks. By comparing the phase diagram for the number of peaks of the probability distribution (Figure \ref{Figure8} $B$) to the phase diagram of the coupling constant (Figure \ref{Figure8} $D$) for the same system with two symmetric genes, we see that larger coupling constants correspond to regions of phase space where there are three peaks. Larger coupling constants result from larger average numbers of protein molecules, which arise from larger production rates yielding a clear separation of the peaks of the probability distributions in $n$ space.

Artificially increasing the number of copies of genes, has been considered as a way to decrease noise in the system \cite{Suel,hasty2}. As we can see from the example presented in Figure \ref{Figure8} $A$, things may not always be so simple. Increasing the number of gene copies, can even result in increasing the number of observed states, if the genes are in the nonadiabatic limit. On the other hand, if the genes are in the adiabatic limit, the genes are already equilibrated, and increasing their copy number should not change the statistics of the probability distribution. However, if increasing the number of gene copies, increases the number of proteins, the system may actually be moved from the nonadiabatic to the adiabatic regime. In such a case increasing the number of copies of the gene changes the observed states and decreases the noise. Here, we considered systems with one type of protein molecules coupling all the genes. If the genes are regulated by an external transcription factor, and there is no self-regulation, the number of occupancy states should not be increased by introducing new copies of the gene and the system will become less noisy. Much as for the case of interacting genes with weak and strong promoters, a nonadiabatic two gene system can also evolve from two adiabatic genes when an organism duplicates its genome. In a system with a high synthesis rate in the enhanced production state, where one gene alone sustains the proteomic reservoir needed by the cell, the binding constant for activator proteins can descrease on evolutionary timescales, leading to a nonadiabatic system. If the binding constants of only one of the genes decreases, the result is a mixed system having one adiabatic and one nonadiabatic gene.

The adiabaticity parameter $\kappa=\frac{h^b n^2_{eff}}{k}$ (where $h^b$ is the binding rate of transcription factors and $k$ is the degradation rate) can be rewritten in terms of the effective number of synthesized molecules $n_{eff}=\frac{g_{on}}{k}$, the equilibrium parameter $n^{\dagger}$ and the unbinding rate $f$ as $\kappa=\frac{f}{k} (\frac{n_{eff}}{n^{\dagger}})^2$. Keeping the adiabaticity parameter $\kappa$ fixed to a small number while scanning the parameter range in the synthesis rate  ($\sim n_{eff}$) and equilibrium parameter $n^{\dagger}$ corresponds to decreasing the unbinding and binding rates while either the synthesis rate or $n^{\dagger}$ increases. Such a procedure corresponds to introducing mutations into the promoter sequence as the production rates are increased. The genes become nonfunctional and cannot be activated for high enough production rates and high enough threshold values. The binding rates become so small that even for a large production rate, the gene will always remain off. On the other hand if the number of proteins in the system increases, for example by increasing production rates, and the binding and unbinding constants do not change as the system becomes more adiabatic, since the binding rate depends on the number of proteins.

\begin{figure}
\includegraphics{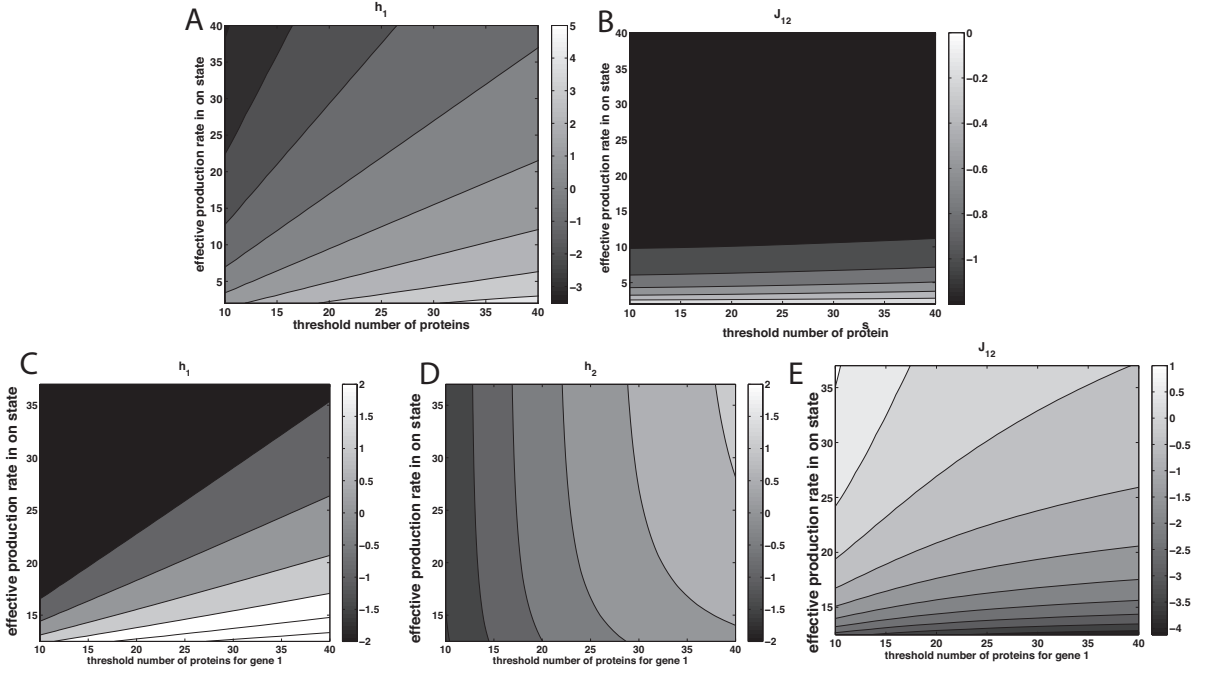}
\caption{The gene $h$ fields for the genes ({\textbf{A}}) and the couplings ({\textbf{B}}) as a function of the effective production rate of the system in the $on$ state $\frac{g_{on}(1)+g_{on}(2)}{k}$ and the threshold number of protein molecules $n^{\dagger}$  for a system with two symmetric nonadiabatic repressed genes. The parameter of both genes are varied. $k=2$, $g_{off}(2)=2$,$\kappa=0.01$. The gene $h$ fields for the repressed gene $1$ ({\textbf{C}}) and the activated gene $2$  ({\textbf{D}}) and couplings ({\textbf{E}}) as a function of the effective production rate of the system gene $1$ in the $on$ state $\frac{g_{on}(1)+g_{on}(2)}{k}$ and the threshold number of protein molecules $n^{\dagger}_1$ of gene $1$ for a system with one nonadiabatic repressed and one nonadiabatic activated gene. The parameters of the repressed gene (gene $1$)are varied, while those of the activated gene (gene $2$) are fixed. The parameters are: $g_{on}(2)=24, g_{off}(2)=2, n^+_1=8$ and $k=2$, $g_{off}(1)=2$.}
\label{Figure14}
\end{figure}

\textbf{Two repressed genes}

Differences between systems having only activated genes and those having repressed genes are also apparent in the nonadiabatic limit. The coupling still is a function mainly of the production rate as can be seen from Figure \ref{Figure14} $A, B$. Unlike for the two activated gene system, the $h$ gene field for the two repressed gene system is also mainly a function of the effective production rate. Such a dependence of the gene fields on the  effective production rate is reminiscent of the behaviour seen in the adiabatic limit, even though the probability distribution still is bimodal and fully nonadiabatic. 

When there is a small average number of protein molecules in the system, a gene which can be repressed is in the $on$ occupancy state, producing large number of proteins. Hence, if the production rate of the gene is large enough, even for genes with slow binding and unbinding constants, the concentration of proteins will become large enough to repress the gene. If the threshold number of protein molecules for switching is large enough, even a nonadiabatic gene, with a very slow binding constant, will become repressed for large effective production rates. However, once a gene is repressed, it will remain repressed, because the unbinding process is slow and its rate does not depend on protein concetration. For this reason, the gene fields for a pair of repressed genes in the nonadiabatic limit behave in a similar way to the gene fields for adiabatic activated genes: the fields decrease with the increase of the effective production rate and increase with the increase of the threshold number of protein molecules. 

For repressed gene systems, the number of protein molecules does not grow with the production rate - it is limited for large production rates. Much as in the case of two activated genes, peaks associated individual expression states of the two gene system can be discerned in the protein number probability distributions.

\textbf{One repressed and one activated gene}

The nonsymmetric nonadiabatic repressed-activated gene system displays the properties of the couplings and fields for genes in the nonadiabatic limit, which are shared by both systems with only activated and only repressed genes (Figure \ref{Figure14} $C, D, E$): the activated gene $h$ fields depend only on the threshold number of protein molecules, whereas the repressed gene $h$ fields depend also on the number of protein molecules in the system. The couplings depend on the number of proteins in the system. As in the adiabatic limit, the couplings counterbalance the repressed gene and activated gene $h$ fields, stabilizing the $on, off$ and $off, on$ states.

In the nonadiabatic limit for a two gene system having a single repressed and a single activated gene, we again can discern peaks in the probability distribution due to individual expression states. As in previously discussed nonadiabatic systems, the multimodality of the probability distribution shows the inadequacy of using the mean and the variance to characterize the steady states. Hence in this case we cannot link the couplings specifically to the noise parameters in as straightforward manner as was possible in the adiabatic limit. For intermediate values of the adiabaticity parameter we find the probability distributions have larger variance than a single effective gene would have. The underlying gene expression states merge to form one pronounced peak for large adiabaticity parameters.


\begin{figure}[h!]
\includegraphics{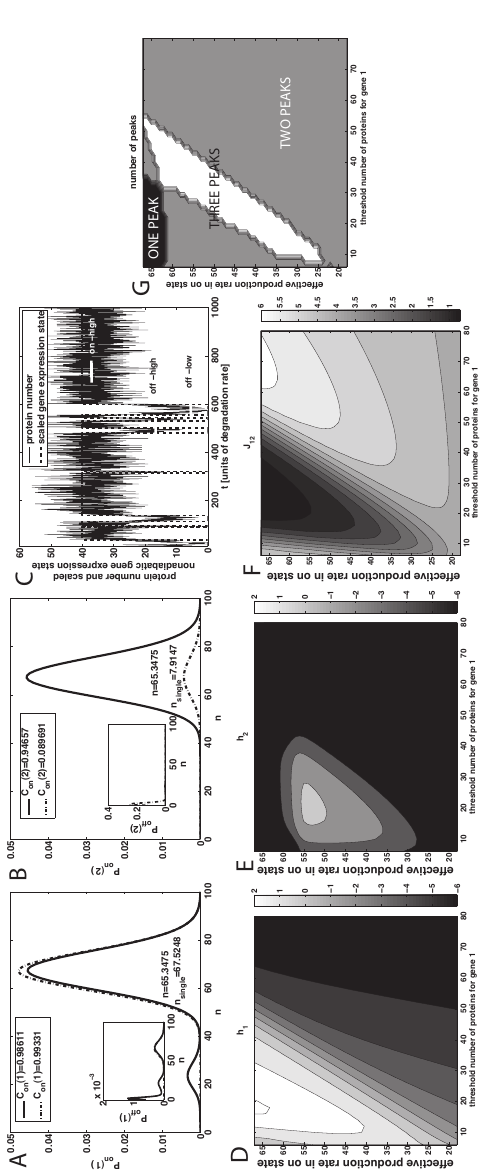}
\caption{}
\label{Figure10}
\end{figure}

\textbf{Fig 8}. {The probability distributions of the number of protein molecules and the gene to be found in the $on$ state for an adiabatic gene $1$ (\textbf{A}) and a nonadiabatic gene $2$ (\textbf{B}) in a system with one adiabatic and one nonadiabatic activated genes (solid lines) compared to the probability distributions of an effective single gene with the same effective production rates as the two gene system. $C_{on}(i)=\sum_n P_{on}(i)$ is the probability of gene $i$ to be in the $on$ state.  $g_{on}(1)=14, g_{off}(1)=1,g_{on}(2)=24, g_{off}(2)=1, n^+_1=6, n^+_2=6,\kappa_1=200, \kappa_2=0.2$ and $k=1$.  \textbf{C}: The number of protein molecules (solid line) as a function of time and the gene expression state of the nonadiabatic gene (dashed line). The gene expression state is scaled to be visible on the same figure as the number of protein molecules. Results of Gillespie simulations of a a mixed adiabatic-nonadiabatic system with two activated genes. $g_{on}(1)=20, g_{off}(1)=1,g_{on}(2)=10, g_{off}(2)=1, n^+_1=8, n^+_2=8,\kappa_1=10000, \kappa_2=0.5$ and $k=1$. The gene $h$ fields for the adiabatic gene $1$ ({\textbf{D}}) and the nonadiabatic gene $2$ ({\textbf{E}}), the couplings ({\textbf{F}}) and the number of peaks of the probability distribution ({\textbf{G}})  as a function of the effective production rate of the system in the $on$ state $\frac{g_{on}(1)+g_{on}(2)}{k}$ and the threshold number of protein molecules $n^{\dagger}_1$ of gene $1$ for a system with one adiabatic and one nonadiabatic activated gene. The parameters for the adiabatic gene (gene $1$) are varied, while those for the nonadiabatic gene (gene $2$) are fixed. The parameters are: $g_{on}(2)=17, g_{off}(2)=1,n^+_2=18, \kappa_1=10000, \kappa_2=5 \cdot 10^{-5}$ }

\textbf{The mixed adiabatic-nonadiabtic activated-activated gene case}

In the case of a mixed system having one adiabatic gene and one nonadiabatic gene changing the state of the nonadiabatic gene modifies the proteomic field. The effective synthesis rate of the adiabatic gene changes and a new steady state appears with intermediate protein molecule number. 

Let us consider a specific case of two activated genes: the expression state of one gene changes on fast timescales (the binding and unbinding reactions are in equilibrium in the presented model), the gene occupancy of the other gene is not rapidly equilibrated. We may say, the first gene has a high adiabaticity parameter, $\kappa$, while the second gene is in the nonadiabatic limit, with a small $\kappa$. Since the probability of binding a transcription factor per unit time depends on the number of protein molecules in the system, we will use a conservative estimate of the adiabaticity parameter of the two gene system. We estimate the mean number of protein molecules, by the mean number of protein molecules if both genes were always on: $\kappa=\frac{h^b}{2k} (\frac{g_{on}(1)+g_{on}(2)}{k})^2$. The effective adiabaticity of a given gene, is therefore in fact a function of the parameters of the system taken as a whole, not merely a function of the parameters of that given gene alone. It is straightforward to note that introducing an adiabatic gene into a system with a functioning nonadiabatic gene, will increase the effective adiabaticity parameter of the previously nonadiabatic gene, because the number of proteins increases. Such a scenario is similar to the case of adding a new gene to a functioning system, without modifying the effective production rates, as described earlier for the case of abundant resources. Therefore a  gene with slow binding parameters, which is surrounded by a constant high concentration of transcription factors, may behave adiabatically when coupled with an adiabatic gene; it is effectively buffered. The gene with slow binding and unbinding alone would be nonadiabatic -- the slow binding and unbinding would not be enough to sustain the constant proteomic cloud and the protein numbers would equilibrate in the two gene states. However, if the large $\kappa$, adiabatic gene has a large enough production rate to sustain a constant proteomic environment, it will push the nonadiabatic gene into the adiabatic regime. If the strictly adiabatic gene does not produce enough protein molecules the nonadiabatic gene will remain in the nonadiabatic limit. 

Even if the interaction with the adiabatic gene does not move the independently nonadiabatic gene into the adiabatic regime, the increase in the number of proteins, results in differences in the probability distributions for systems containing both genes compared to what would be found for an effective single gene systems (with the same effective production rate and adiabaticity as the two gene system). Since the adiabatic gene provides a constant reservoir of high numbers of protein molecules, this leads to the appearance of an additional peak in the probability distribution (Figure \ref{Figure10} $A$). The new peak corresponds to the nonadiabatic gene being in the $off$ state and the adiabatic gene producing proteins at the equilibrated rate. The mean expression rate of the adiabatic gene is affected by the production rate of the nonadiabatic gene, through the number of protein molecules present in the system. The original peak of the adiabatic distribution remains, corresponding to the nonadiabatic gene being in the $on$ state. The third peak corresponds to a state when the nonadiabatic gene is $off$ and the number of protein molecules has decreased to effectively deactivate the adiabatic gene. Hence, the effective production rate of the adiabatic gene is modified by the state of the nonadiabatic gene. The low and high, effective production rate states of the adiabatic gene are an emergent form of bistability induced by the change in the average number of protein molecules due to different occupancy states of the nonadiabatic gene (Figure \ref{Figure10} $B$). Experimentally we would observe three steady states. If proteins produced by one gene could be distinguished from those produced by the other gene, for example by coexpressing different coloured fluorophores, one could measure the two different time scales for switching: one which results from a rare change in the gene expression state and the other from a change in the available number of protein molecules. Hence in the case of an mixed adiabatic-nonadiabtic system, the cooperative steady state imposes additional states that may be observed focussing only on the adiabatic gene. An extract from a simulation trajectory for such a mixed adiabatic-nonadiabtic system is presented in Figure \ref{Figure10} $C$. We can clearly distinguish the three states. We can also see how the nonadiabatic gene state changes on slow timescales. In the discussed example the adiabatic gene states are equilibrated and switch on fast timescales (not shown).

If the production rate from the nonadiabatic gene is smaller than that for the adiabatic gene, so that the mean number of protein molecules of the two gene system does not significantly change compared to that for the effective adiabatic system, the intermediate peak, due to the $off$ nonadiabatic state and the high production adiabatic state may be indistinguishable from the $on-high$ nonadiabatic-adiabatic state. In this case the probability of the nonadiabatic gene to be in the $off$ state is very small, as can be see from the small positive $h_2$ gene fields in Figure \ref{Figure10} $D, E, F$. Figure \ref{Figure10} $G$  also displays a phase diagram for the observed number of peaks of the probability distribution, in which we vary the parameter of the adiabatic gene. Comparing the phase diagram for the coupling parameter to the phase diagram for the number of peaks, we see that three peak situation appears for relatively small couplings, when the adiabatic gene has an intermediate tendency to be $on$, but where this tendency is large enough so that the adiabatic gene can sustain its own proteomic field.  If the adiabatic gene has a high equilibrium constant, the adiabatic gene produces small protein numbers and the nonadiabatic $off$-adiabatic protein numbers overlap with the state when both genes are $off$. Once again we note the adiabatic and nonadiabatic characteristics in the gene $h$ fields and couplings. When the parameters of the nonadiabatic gene are varied the coupling phase diagram changes only with the effective production rate of the system (results not shown).  


\begin{figure}
\includegraphics{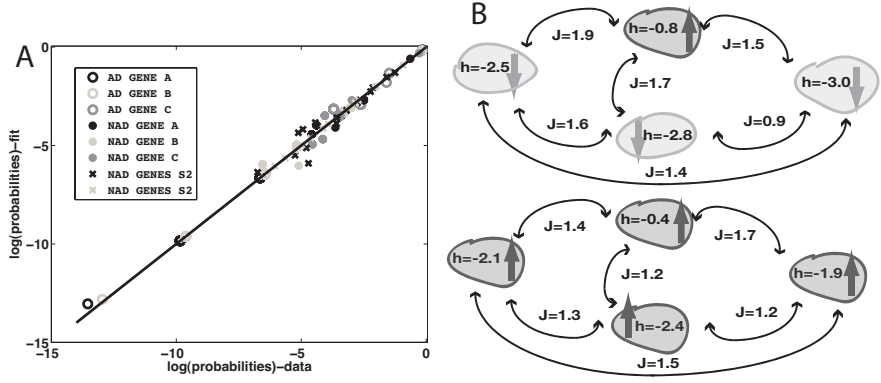}
\caption{{\textbf{A}: Comparison of a maximum entropy model fit to probabilities of all sixteen possible gene occupancy states for a four gene system interacting by one type of protein. Different symbols indicate genes with different parameters. Gene A: $g_{on}(1)=24, g_{off}(1)=2$,$n^+=5$. Gene B: $g_{on}(1)=12, g_{off}(1)=1$,$n^+=3$.  Gene C: $g_{on}(1)=17, g_{off}(1)=1$,$n^+=8$. $k=2$ for all systems. $S1=\{ A,B,C,g_{on}(1)=12, g_{off}(1)=1,n^+=10\}$. $S2=\{ A,B,C,g_{on}(1)=12, g_{off}(1)=1,n^+=5\}$}. \textbf{B}:  A schematic representation of the mapping of the $S1$ and $S2$ systems.}
\label{Fig_probfit}
\end{figure}

\subsection*{Generalization to many genes}
In many composite systems introducing effective two-body interactions between the constituents is able to account for most of the behaviour of many body systems having a large number of parts. To see whether this is true for these networks of interacting gene switches we examine networks of four interacting genes and test whether the results from a complete treatment are consistent with those coming from the Ising mapping based on two body interactions. We fit the Ising model parameters using a minimally constrained maximum entropy model. Within a parametrized two body interaction model, the maximum entropy approach corresponds to minimizing the Kullback-Leibler divergence between two- point correlation functions. We consider examples of genes, both in the adiabatic and nonadiabatic regime and with varying coupling constants and internal $h$ fields. The probabilities predicted based on the fitted Ising model are compared to those from simulations in Figure \ref{Fig_probfit} A. The predicted three and four point correlations based on the coupling constant and random fields predicted from two body correlations agree very well with the correlations derived from the data. The correlations decrease as the probabilities of the particular states become smaller and more uniform between the substates. As the probabilities of each of the substates become closer to each other the couplings increase. The mapping is generally successful. It works well for genes that have large and small coupling constants, both in the adiabatic and nonadiabatic regime.

The fitted couplings and fields for the four gene system agree with the intuition we have gained based on the two gene system. A gene with a positive gene $h_i$ field has a higher tendency to be in the $on$ state than in the $off$ state. For identical genes, the coupling constant between two genes plays a role when the probability of the genes to be $on$ is close to $0.5$. Since all studied genes in the four gene system are activated genes, all the coupling constants are positive. The couplings properly describe in a quantitative way the collective stabilization of the proteomic field by all the genes. 

In the examples presented, the genes in the system with identical genes $A$ and $B$ are more likely to be found in the $on$ state ($C_{on}(i)>0.5$), both in the adiabatic and nonadiabtic limit. In the adiabatic limit the genes are mapped onto an Ising system with positive fields and small couplings. In the nonadiabatic limit, the fields are negative. Guided by the results from studying a two gene system, we see that each gene alone could not sustain the $on$ state if the other genes were constrained to be in the $off$ state. Large coupling constants, which are related to the number of proteins in the system, result in the genes being more likely to be $on$ than $off$. The genes in the system with identical genes $C$ is more likely to be found in the $off$ state, both in the adiabatic and nonadiabtic limit. In both cases the $C$ genes are described by negative $h$ fields. Since the probability of the genes to be $on$ is close to a half the coupling constants are of the same order of magnitude as the $h$ fields. In the examples with four different types of genes in the nonadiabatic limit, the genes all have negative $h$ fields. In the $S1$ system, only one gene has a higher probability to be $on$ than to be $off$. That gene has the smallest magnitude of the $h\approx -0.75$ field compared to other genes with $-3.05<h<-2.5$. The couplings have values between $J \approx 0.9$, between the two genes which are least probable to be $on$, and $J \approx 1.9$, between the two genes which are most probable to be $on$. Pairs of genes with higher probabilities to be $on$ contribute more to the common proteomic pool and hence have stronger couplings. In the $S2$ system, the threshold number of proteins of one of the genes is decreased. The number of proteins in the system increases compared to the $S1$ case and all genes are now more probable to be $on$. Only the $h$ gene field of the gene the chemical parameters of which were varied, changed significantly (from $-3.05$ in the $S1$ system to $-1.95$ in $S2$). The coupling constants of that gene with the other genes increases, whereas the other coupling constants decrease (Figure \ref{Fig_probfit} $B$).


A Boolean representation is able to capture the properties of a genetic network. It is worth pointing out, that such a Boolean approximation is ``tailor made" for the present sort of system \cite{guetleibler}. The parameters of the mapping are based on a detailed molecular description in this specific parameter regime. Changing the parameters of the system will change the probability distributions and the corresponding fit parameters of the corresponding spin system. As we could see from the two gene discussion the dependence of the Ising model parameters on the chemical parameters is far from trivial. However, the observation that two point correlations account for a large part of the correlations in highly coupled many gene systems is worth exploiting in practical models. We believe the mapping onto an effective spin system that we have described here can facilitate the study of larger networks and be used to explore the existence of multiple steady states and characterize them. 

For the systems where all four genes are identical, the gene fields $h$ and coupling $J_{ij} $ are identical for all genes. Such systems composed of identical genes are especially straightforward to generalize to larger systems. The values of the fields and couplings are also the same for systems with four and two genes, if the steady state probabilities of the genes in both cases are the same. Hence finding the parameters of the model for a small network allows us to study larger networks repetitively made up of these smaller subsystems. The present approach can be easily generalized to systems having many different proteins.

\begin{figure}
\includegraphics{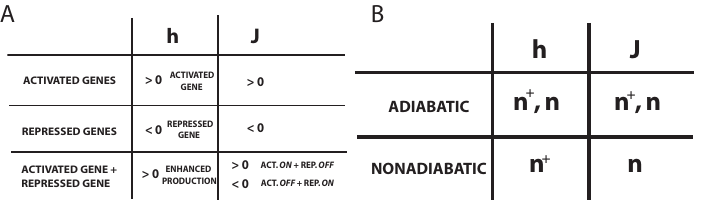}
\caption{\textbf{A}: A summary of the values of the $h$ fields and coupling constants $J$ for interacting two activated, two repressed and one activated- one repressed gene systems. \textbf{B}: A summary of the dependence of the $h$ fields and coupling constants $J$ on the number of protein molecules in the system $n$ and the threshold $n^{\dagger}$ in the adiabatic and nonadiabatic regime.}
\label{diag}
\end{figure}


\subsection*{Conclusions}
Gene networks consist of interacting genes, the protein molecules that are produced in response to the gene state and that can, in turn, bind to different genes. A molecular kinetic description of the underlying interactions is important to capture the properties of these genetic regulatory systems. We considered a number of self-interacting small gene regulatory systems. We showed how to find reduced descriptions of such systems using gene state variables alone. Figure \ref{diag} A shows a summary of the values of the $h$ fields and coupling constants for interacting activated and repressed genes. In Figure \ref{diag} B we also summarize the primary ways in which the parameters of the reduced model depend on both the total number of protein molecules in the system and the threshold for switching in the adiabatic and nonadiabatic regime. Although within a deterministic description, genes with identical binding parameters can generally be reduced to an effective one gene model in the adiabatic limit (where gene expression states can be taken to be in equilibrium), the present analysis shows that the genes cannot be treated as independent. The coupling between genes results in slightly different steady state characteristics and a much sharper transition from the $on$ to $off$ state. Furthermore a composite one gene description does not reproduce the steady state properties of even adiabatic genes with different equilibrium constants. In all cases of considered systems, the steady state is a product of a cooperative mutual proteomic field, which is both produced by and used by both genes. 

In the nonadiabatic limit, when protein molecule numbers can come to steady states faster than the gene occupancy states can change, multipeaked probability distributions for the number of protein molecules emerge. These corresponding to protein molecule numbers equilibrating in specific multiple gene expression states. For a mixed adiabatic-nonadiabatic two activated gene system, the gene expression state of the nonadiabatic gene can modify the proteomic field and induce a change in the expression state of the adiabatic gene. 

If the number of protein molecules increases, but the binding and unbinding rate coefficients do not change, the binding rate of the system increases and the nonadiabatic system can become adiabatic. In this case the noise in the system can be reduced by introducing more copies of existing genes. If the binding rates of a single gene are in the adiabatic limit, however, increasing the number of genes will not reduce the noise coming from gene occupancy state fluctuations, but can reduce the noise from the birth and death of proteins in small numbers. 

Even when the systems we have studied are approximated by an effective gene description, the protein numbers are not generally well described by a Poisson distribution. If the probability distribution of the number of protein molecules were described  properly by a Poisson distribution, the deterministic kinetic rate equations would follow. For unimodal probability distributions we can link the deviation from a Poisson description to a nonzero coupling constant of an effective pseudo-Ising spin system. For a system with a bistable solution the same nonzero coupling informs us of the bimodality of the distribution. In both cases, if the couplings are small or vanishing, this would indicate that description in terms of independent genes is valid. Such a description works in those parameter regimes, where genes can individually sustain the proteomic field needed for the same steady state as is observed in the two gene system. For pure repressed and activated gene systems this is possible when each gene is expressed at maximum rates. In the case of a mixed activated-repressed gene system, the genes do not cooperate when each gene on average produces similar numbers of protein molecules.

The effective description of a stochastic interacting protein-gene network in terms of Ising like variables indicating the occupancy and therefore expression states of genes, is a first step to deriving a reduced molecular based description of large networks of many genes. The mapping developed here shows how activation is linked to positive couplings and repression to negative couplings. The gene $h$ fields describe the genes preference to be $on$ regardless of the state of the other genes. These fields are local and depend on the chemical kinetic parameters of the genes. The fields can have different signs for different genes in the same system. 

In the nonadiabatic limit we have shown how correlations provide a clear signature of the lack of equilibration of gene expression states, by showing the dependence of the couplings mainly on protein molecule numbers and of the local gene $h$ fields mainly on threshold protein molecule numbers. 

The proposed mapping of a molecular kinetic  description of gene networks onto a Boolean description should allow the study of the dynamics of larger systems of genes within this reduced framework. We hope this approach can aid simulations in the nonadiabatic limit which are computationally expensive due to the slow binding rates. Initially, each system must be mapped onto this effective description.

We considered extensions of the activated gene systems to larger small networks using maximum entropy techniques and found that the appropriate Boolean description describes correctly predicts higher order correlations in the composite gene system. This exploratory study suggests two body interactions can probably be used quite effectively to capture the properties of larger networks.

\subsection*{Acknowledgements}
We are grateful for helpful discussions with and helpful comments of Anat Burger. The work was supported by NSF grant PHY0822283  to the Center for Theoretical Biological Physics. 

\subsection*{Appendix A}
Consider two genes coupled by a common protein environment of activators numbering $n$. The master equation for the evolution of the four states of the system is $P_{ij} =\{P_{on,on}(n),P_{off,on}(n),P_{on,off}(n),P_{off,off}(n) \}$:
 \begin{eqnarray*}
\frac{\partial P_{on, on}(n)}{\partial t}&=&(g^{on}_{1}+g^{on}_{2})[P_{on, on}(n-1)-P_{on, on}(n)]+k [(n+1)P_{on, on}(n+1)-nP_{on, on}(n)]+
\nonumber \\
&+& \frac{h^b_1}{2} n(n-1) P_{off, on}(n)+ \frac{h^b_2}{2} n(n-1) P_{on, off}(n)-(f_1+f_2)P_{on, on}(n)\nonumber \\
\frac{\partial P_{on, off}(n)}{\partial t}&=&(g^{on}_{1}+g^{off}_{2})[P_{on,  off}(n-1)-P_{on,  off}(n)]+k [(n+1)P_{on,  off}(n+1)-nP_{on, off}(n)]+
\nonumber \\
&+& \frac{h^b_1}{2} n(n-1) P_{off, off}(n) +f_2 P_{on, on}(n)-(\frac{h^b_2}{2} n(n-1)+f_1) P_{on, off}(n)\nonumber \\
\frac{\partial P_{off, on}(n)}{\partial t}&=&(g^{off}_{1}+g^{on}_{2})[P_{off, on}(n-1)-P_{off, on}(n)]+k [(n+1)P_{off, on}(n+1)-nP_{off, on}(n)]+
\nonumber \\
&+& \frac{h^b_2}{2} n(n-1) P_{off, off}(n)+ f_1 P_{on, on}(n)-(\frac{h^b_1}{2} n(n-1)+f_2) P_{off, on}(n)\nonumber \\
\frac{\partial P_{off, off}(n)}{\partial t}&=&(g^{off}_{1}+g^{off}_{2})[P_{off, off}(n-1)-P_{off, off}(n)]+k [(n+1)P_{off, off}(n+1)-nP_{off, off}(n)]+
\nonumber \\
&+& f_1 P_{on, off}(n)+ f_2 P_{off, on}(n)-(\frac{h^b_1}{2} n(n-1)+\frac{h^b_2}{2} n(n-1)) P_{off, off}(n)\nonumber \\
\end{eqnarray*}

\subsection*{Appendix B}
In this appendix we comment on the choice of $s_i=0,1$ in making the magnetic analogy, as opposed to the symmetric choice $s_i=-1,+1$. In the latter case, for the two gene case, the explicit mapping becomes:
\begin{eqnarray}
\tilde{h}_1=\frac{1}{4}\log \frac{C_{on, on} C_{on, off}}{C_{off, on}C_{off, off}}\nonumber \\
\tilde{h}_2=\frac{1}{4}\log \frac{C_{on, on} C_{off, on}}{C_{on, off}C_{off, off}}
\label{hitwo1}
\end{eqnarray}
\beq
\tilde{J}_{12}=\frac{1}{4}\log \frac{C_{on, on} C_{off, off}}{C_{off, on}C_{on, off}}
\label{hitwo2}
\eeq
Positive values of the gene fields describe the preference of the gene to be in the on state, regardless of the state of the other gene. The gene fields in the nonsymmetric model (Eq. \ref{hitwo2}) describe the genes own preference to be in the on state, assuming the other gene is in the $off$ state, so that the other gene does not contribute to the common protein pool. In the symmetric model (Eq. \ref{hitwo1}), the interaction due to sharing a common field is incoorporated into the preference of the gene. The interaction term, although it has the same form as in the nonsymmetric case (compare Eqns \ref{hitwoU2} and \ref{hitwo2}), up to constant multiplicative factor, describes how a state with both genes in the same expression state is stabilized. The smaller value of the interaction in the symmetric model compared to the nonsymmetric results from encompassing parts of the protein mediated $J_{ij}$ into the gene fields. The reclassification of the interaction terms becomes especially clear in the nonadiabatic limit. Unlike in the symmetric model, where the gene fields depends mainly on the genes affinity for binding a protein, for the symmetric model, the gene fields show a dependence on protein concentrations. Both the symmetric and nonsymmetric model predict that two-body interactions are sufficient to describe many gene interactions.

\end{document}